\documentclass[a4paper,10pt,conference,romanappendices]{IEEEtran}
\IEEEoverridecommandlockouts
\usepackage{algorithm,algpseudocode,amsmath,amssymb,amsthm,bbm,cite,color,enumitem,graphicx,microtype,setspace,url}
\usepackage[USenglish]{babel}
\usepackage{hyperref}
\usepackage[utf8]{inputenc}
\usepackage[T1]{fontenc}
\usepackage{mathtools}
\usepackage{rotating}
\usepackage{cuted}
\usepackage{epsfig}
\usepackage{float}
\usepackage{subfigure}
\usepackage{subcaption}
\usepackage{anyfontsize}
\usepackage{t1enc}
\usepackage{cite}
\usepackage{textcomp}
\usepackage{latexsym}
\usepackage{multirow,multicol}
\usepackage{tabularx,booktabs}
\usepackage{moreverb}

\usepackage{tikz,circuitikz,pgfplots}
\pgfplotsset{compat=1.17}
\usetikzlibrary{arrows, arrows.meta, backgrounds, calc, intersections, decorations.markings, decorations.pathreplacing, positioning, shapes}
\pgfplotsset{compat=newest}
\pgfplotsset{
/pgfplots/ybar legend/.style={
/pgfplots/legend image code/.code={%
\draw[##1,/tikz/.cd,bar width=0.1cm,yshift=-0.2em,bar shift=0.5*\pgfplotbarwidth]
plot coordinates {(0.5*\pgfplotbarwidth,0.6em) (2.5*\pgfplotbarwidth,0.4em) (4.5*\pgfplotbarwidth,0.2em)};},}}

\usepackage[figurename=Fig., labelsep=period, font=small]{caption}

\newcommand{\e}{\mathbf{e}}

\newcommand{\h}{\mathbf{h}}

\newcommand{\p}{\mathbf{p}}

\newcommand{\x}{\mathbf{x}}
\newcommand{\y}{\mathbf{y}}
\newcommand{\z}{\mathbf{z}}

\newcommand{\0}{\mathbf{0}}

\newcommand{\E}{\mathbf{E}}

\newcommand{\G}{\mathbf{G}}
\renewcommand{\H}{\mathbf{H}}
\newcommand{\I}{\mathbf{I}}

\renewcommand{\P}{\mathbf{P}}

\newcommand{\W}{\mathbf{W}}

\newcommand{\Y}{\mathbf{Y}}
\newcommand{\Z}{\mathbf{Z}}

\newcommand{\etab}{\boldsymbol{\eta}}

\newcommand{\setC}{\mathcal{C}}

\newcommand{\setN}{\mathcal{N}}

\newcommand{\Compl}{\mbox{$\mathbb{C}$}}

\newcommand{\blkdiag}{\mathrm{blkdiag}}

\newcommand{\herm}{\mathrm{H}}

\newcommand{\sgn}{\mathrm{sgn}}

\newcommand{\tran}{\mathrm{T}}

\definecolor{oulu_blue}{HTML}{23408F}
\definecolor{oulu_green}{HTML}{39B54A}

\definecolor{red}{rgb}{1,0,0}
\definecolor{red_magenta}{rgb}{1,0,0.5}
\definecolor{magenta}{rgb}{1,0,1}
\definecolor{blue_magenta}{rgb}{0.5,0,1}
\definecolor{blue}{rgb}{0,0,1}
\definecolor{blue_cyan}{rgb}{0,0.5,1}
\definecolor{cyan}{rgb}{0,1,1}
\definecolor{green_cyan}{rgb}{0,1,0.5}
\definecolor{green}{rgb}{0,1,0}
\definecolor{green_yellow}{rgb}{0.5,1,0}
\definecolor{yellow}{rgb}{1,1,0}
\definecolor{red_yellow}{rgb}{1,0.5,0}

\ifCLASSINFOpdf
\else
\fi

\begin{document}

\title{Sparse Near-Field Channel Estimation for XL-MIMO via Adaptive Filtering}
\author{
\IEEEauthorblockN{Vidya Bhasker Shukla and Italo Atzeni}
\IEEEauthorblockA{Centre for Wireless Communications, University of Oulu, Finland
\\E-mail: \{bhasker.shukla, italo.atzeni\}@oulu.fi}
\thanks{This work was supported by the Research Council of Finland (336449 Profi6, 348396 HIGH-6G, 357504 EETCAMD, and 369116 6G~Flagship).}}
\maketitle

\begin{abstract}
%The integration of extremely large-scale multiple-input multiple-output (XL-MIMO) systems with sub-terahertz (sub-THz) frequencies is necessary to meet the key performance indicators of next-generation wireless communication. However, with the evolution of  XL-MIMO, currently ongoing wireless communication is progressively shifting from the far-field to the near-field (NF) domain. Accurate modeling of the NF channel requires adopting the spherical wavefront assumption, which significantly complicates the channel estimation process. In the context of NF channel estimation, we have proposed polar-domain zero-attracting least mean squares (PD-ZALMS) algorithm to estimate the NF sub-THz XL-MIMO channel. Simulation results show that the PD-ZALMS algorithm offers superior estimation accuracy compared to conventional methods and even the oracle least-squares estimator despite its lower computational complexity and less pilot overhead requirements.
Extremely large-scale multiple-input multiple-output (XL-MIMO) systems operating at sub-THz carrier frequencies represent a promising solution to meet the demands of next-generation wireless applications. This work focuses on sparse channel estimation for XL-MIMO systems operating in the near-field (NF) regime. Assuming a practical subarray-based architecture, we develop a NF channel estimation framework based on adaptive filtering, referred to as \textit{polar-domain zero-attracting least mean squares (PD-ZALMS)}. The proposed method achieves significantly superior channel estimation accuracy and lower computational complexity compared with the well-established polar-domain orthogonal matching pursuit. In addition, the proposed PD-ZALMS is shown to outperform the oracle least-squares channel estimator at low-to-moderate signal-to-noise ratio.
\end{abstract}

\section{Introduction}
To support the growing demands of next-generation wireless applications, there is an increasing need for ultra-high data rates and sensing resolution~\cite{Raj20}. Meeting these requirements involves increasing the carrier frequencies up to the sub-THz range to unlock wider bandwidths~\cite{wang2023road}, leveraging reconfigurable intelligent surfaces (RISs) to dynamically control the propagation environment~\cite{ding2022state}, and employing massive multiple-input multiple-output (MIMO) arrays to enable precise beamforming and extensive spatial multiplexing.
%The true potential of these technologies lies in their ability to integrate and complement each other effectively.
Furthermore, extremely large-scale (XL) array technologies, such as XL-RIS~\cite{yang2023channel} and XL-MIMO~\cite{cui2022near}, deployed in the sub-THz band are especially promising for realizing advanced communications and sensing capabilities.

By employing hundreds or even thousands of antennas, XL-MIMO allows to generate highly focused beam patterns, enhancing spatial resolution and beamforming gain. Due to the large array aperture and the high carrier frequencies, XL-MIMO systems typically operate in the near-field (NF) regime. In this regime, the Fraunhofer distance, which defines the boundary between NF and far-field regions, can expand to several hundred meters~\cite{cui2022channel}. As a result, XL-MIMO channel modeling, channel estimation, and beamforming design must account for the spherical nature of the wavefronts, as opposed to the planar-wave assumption characterizing far-field propagation. In this respect,~\cite{cui2022channel} introduced a polar-domain dictionary to represent the NF beamspace, featuring uniform sampling in the angular domain and non-uniform sampling in the distance domain. Additionally,~\cite{elbir2023nba} investigated wideband channel estimation, analyzing the NF beam squint phenomenon and proposing techniques for joint angle-distance estimation. Studies on NF parameter estimation in XL-MIMO have also developed decoupled strategies, which simplify the estimation of angles and distances and thus reduce the computational complexity~\cite{huang2023low}.

All the aforementioned approaches rely on processing pilot signals collected across the entire array. In contrast, subarray-based methods exploit localized processing within smaller array segments to enhance the efficiency of channel estimation~\cite{zhu2023sub}. Subarray-based architectures have recently gained significant attention, especially at sub-THz frequencies~\cite{chen2023hybrid}, as they offer enhanced scalability and design flexibility, simplified circuitry, and reduced signal processing complexity. Studies on beamforming design and spectral efficiency analysis of subarray-based architectures have demonstrated that, with the same number of antennas, enhanced performance can be achieved by increasing the inter-subarray spacing, thereby amplifying NF effects~\cite{yan2021joint}. Motivated by these findings, this work adopts a subarray-based architecture to capitalize on these advantages.

\tikzstyle{arrow} = [thin,->,>=stealth]
  \usetikzlibrary{shapes.geometric}
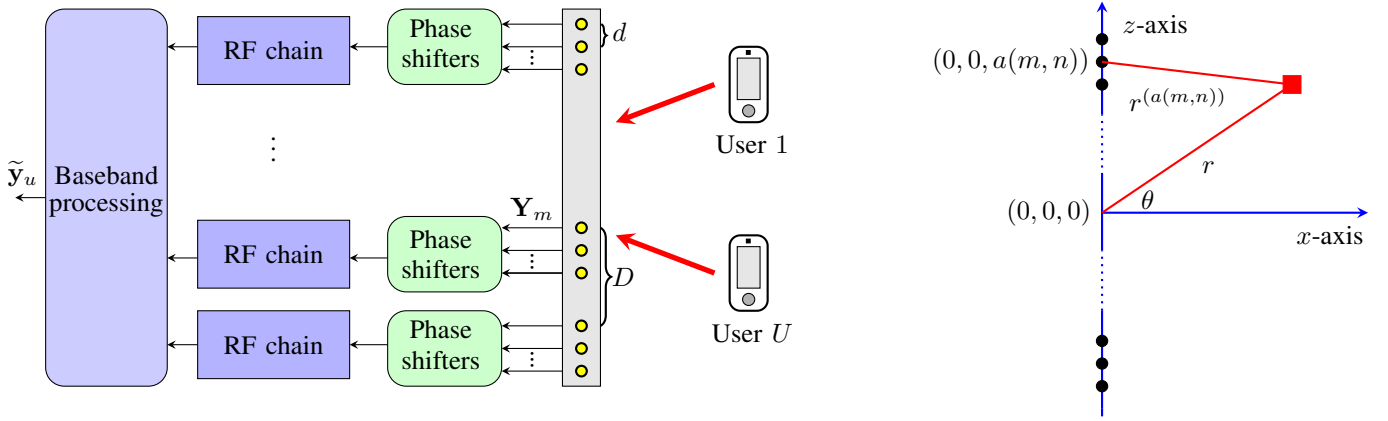
\begin{figure*}	[t!]
	%\centering
	\begin{tikzpicture}	[scale=1, transform shape]
  
		\draw [rounded corners=7pt, fill=blue!20!white]
		(0,-1) rectangle (1.6,4) ;
        %\draw [arrow] (0,3)--(-.4,3);
       % \draw [arrow] (0,2.5)--(-.4,2.5);
        %\node at (-.2,1.8) {$\vdots $};
        \draw [arrow] (0,1.5)--(-.4,1.5);
  %       \draw [rounded corners= 7 pt, fill=green!5!white]
		% (4.5,-1) rectangle (6,4) ; 
		\node at (.8,1.8) {Baseband};
		\node at (.8,1.4) {processing};        
		\draw [rounded corners=7pt, fill=green!20!white](4.5,3) rectangle (6,4);
       	\draw [rounded corners=7pt, fill=green!20!white](4.5,-1) rectangle (6,0);
        \draw [rounded corners=7pt, fill=green!20!white](4.5,.25) rectangle (6,1.25);
        %\node at (5.2,2.3) {$\vdots $};
        \node at (5.2,3.75) {Phase};
		\node at (5.2,3.35) {shifters};
		
	  \node at (5.2,.95) {Phase};
		\node at (5.2,.55) {shifters};	
        
 \node at (5.2,-.25) {Phase};
		\node at (5.2,-.65) {shifters};

       % \node at (4.25,3.9) {$\mathbf{y}_1$};
		
		\draw [rounded corners=0pt, fill=blue!30!white]
		(2,3) rectangle (4,3.90) node [pos=.5]{RF chain};	
        \draw [arrow] (2,3.5)--(1.6,3.5);

         \draw [arrow] (4.5,3.5)--(4,3.5);

\draw [arrow] (2,.7)--(1.6,.7);

 \draw [arrow] (4.5,.7)--(4,.7);
  \draw [arrow] (2,-.45)--(1.6,-.45);  

  \draw [arrow] (4.5,-.45)--(4,-.45);
		%\draw [rounded corners=7pt, fill=blue!30!white](1.35,.25) rectangle (2.9,1.20) node[pos=.5] {DAC};
		\draw [rounded corners=0pt, fill=blue!30!white]
		(2,.3) rectangle (4,1.2) node[pos=.5] {RF chain};
        %\node at (2.1,1.5) {$\widetilde{\y}_{u,m}$};
        \draw [rounded corners=0pt, fill=blue!30!white]
		(2,-.9) rectangle (4,0) node[pos=.5] {RF chain};
		\node at (3,2.25) {$\vdots$};
		%\node at (3,1.7) {$ M $};

        %\node at (7.6,2.0) {$\vdots$};

       % \node at (8.2,3.1) {$ s_{{1,q}} $};
        %\node at (8.2,.5) {$ s_{{U,q}} $};

        \node at (-.3,1.82) {$ \widetilde{\mathbf{y}}_{u} $};
		
		\draw [line width=.5] (6.05,.5)--(6.8,.5);
		%\draw [line width=.5]   (6.3,.5) -- (6.3,.75) ;
		%\node  at (6.3,.75) {$ \blacktriangledown $};	

\draw [rounded corners= 0 pt, fill=gray!20!white]
		(6.8,-1) rectangle (7.3,4) ;
       
\draw [thick,fill=yellow!95!white ] (7.05,3.8) circle (.2em );
\draw [thick,fill=yellow!95!white ] (7.05,3.5) circle (.2em );
\draw [thick,fill=yellow!95!white ] (7.05,3.2) circle (.2em );
\draw [thick,fill=yellow!95!white ] (7.05,1.1) circle (.2em );
\draw [thick,fill=yellow!95!white ] (7.05,.8) circle (.2em );
\draw [thick,fill=yellow!95!white ] (7.05,.5) circle (.2em );
\draw [thick,fill=yellow!95!white ] (7.05,-.2) circle (.2em );
\draw [thick,fill=yellow!95!white ] (7.05,-0.5) circle (.2em );
\draw [thick,fill=yellow!95!white ] (7.05,-.8) circle (.2em );

\usetikzlibrary{decorations.pathreplacing}
  %\draw [thick,decorate,decoration={ brace,amplitude=4pt}]
  \draw [thick,decorate,decoration={brace,amplitude=4pt,mirror}]

        (7.3,-.2) -- (7.3,1.1);
        
\draw [thick,decorate,decoration={brace,amplitude=2pt}]
        (7.31,3.8) -- (7.31,3.5);
    % Label for the bracket
    \node at (7.6,.45) {$D$};
\node at (7.55,3.7) {$d$};
\draw[arrow, thin] (6.8,.5) -- (6,.5);
\draw[arrow, thin] (6.8,.8) -- (6,.8);
\draw[arrow, thin] (6.8,1.1) -- (6,1.1);  

\draw[arrow, thin] (6.8,-.8) -- (6,-.8);
\draw[arrow, thin] (6.8,-.5) -- (6,-.5);
\draw[arrow, thin] (6.8,-.2) -- (6,-.2);  

%\node at (6.5,4.1) {$N$};
\draw[arrow, thin] (6.8,3.8) -- (6,3.8);
\draw[arrow, thin] (6.8,3.5) -- (6,3.5);
\draw[arrow, thin] (6.8,3.2) -- (6,3.2);

\draw[rounded corners=4pt, thick] (9,3.5) rectangle (9.5,2.5);

\node at (9.3,2.2) {User $1$};
% Screen
\draw[fill=gray!20] (9.1,2.8) rectangle (9.4,3.35);

% Home button
\draw[fill=black!30] (9.25,2.65) circle (0.08);

% Speaker slit
\draw[fill=black] (9.22,3.4) rectangle (9.27,3.45);

\draw [ arrow, red, line width = 2pt](8.8,3) -- (7.5,2.5);

\draw[rounded corners=4pt, thick] (9,0) rectangle (9.5,1);

\node at (9.3,-.3) {User $U$};

% Screen
\draw[fill=gray!20] (9.1,.3) rectangle (9.4,.85);

% Home button
\draw[fill=black!30] (9.25,.15) circle (0.08);

% Speaker slit
\draw[fill=black] (9.22,.9) rectangle (9.27,.95);

 \draw [ arrow, red, line width=2pt](8.8,.5) -- (7.5,1);

% \draw [thick,fill=yellow!95!white ] (5.2,3.5) circle (.8 em );

% \draw [ arrow, thick](4.8,3.1) -- ++(.75,.75);

% \draw [thick,fill=yellow!95!white ] (5.2,.7) circle (.8 em );
% \draw [ arrow, thick](4.8,.3) -- ++(.75,.75);

% \draw [thick,fill=yellow!95!white ] (5.2,-.45) circle (.8 em );

 %\draw [ arrow, thick](4.8,-.85) -- ++(.75,.75);

\node at (6.4,3.43) {$\cdot$};
\node at (6.4,3.35) {$\cdot$};
\node at (6.4,3.27) {$\cdot$};

\node at (6.4,1.35) {$\Y_{m}$};

\node at (6.4,.73) {$\cdot$};
\node at (6.4,.65) {$\cdot$};
\node at (6.4,.57) {$\cdot$};

\node at (6.4,-.57) {$\cdot$};
\node at (6.4,-.65) {$\cdot$};
\node at (6.4,-.73) {$\cdot$};

% Right figure (coordinate system)

\hspace{3.9cm}

\draw [blue, thick](10,-1.3) -- (10,0);
\draw [blue, dotted, thick](10,0) -- (10,.8);
\draw [blue, thick](10,.8) -- (10,1.8);
\draw [blue, dotted, thick](10,1.8) -- (10,2.6);
\draw [arrow, blue, thick](10,2.6) -- (10,4.1);

\draw [blue, thick](10,-1.3) -- (10,-1.4);

\draw [ arrow, blue, thick](10,1.3) -- (13.5,1.3);
\node at (13,1) {$x$-axis};

\node[draw,circle,fill,inner sep=1.5pt,] at (10,-1) {};
\node[draw,circle,fill,inner sep=1.5pt,] at (10,-.7) {};
\node[draw,circle,fill,inner sep=1.5pt,] at (10,-.4) {};

\node[draw,circle,fill,inner sep=1.5pt,] at (10,3) {};
\node[draw,circle,fill,inner sep=1.5pt,] at (10,3.3) {};
\node[draw,circle,fill,inner sep=1.5pt,] at (10,3.6) {};

\draw [red, thick](10,1.3) -- ++(2.5,1.7);
\draw [ red, thick](12.5,3) -- (10,3.3);
\node[draw,rectangle,fill,red] at (12.5,3) {};
\node at (11.4,1.9) {$r$};
\node at (10.6,1.47) {$\theta$};
\node at (10.7,3.8) {$z$-axis};
\node at (8.8,3.3) {$(0,0,a(m,n))$};
\node at (11,2.8) {$r^{(a(m,n))}$};
\node at (9.3,1.3) {$(0,0,0)$};
\end{tikzpicture}

\caption{Left: System model; Right: coordinate system.} \label{fig:mmWave_impr1}
\end{figure*}

Existing works on sparse channel estimation for XL-MIMO systems typically employ greedy algorithms from compressive sensing, e.g., based on orthogonal matching pursuit (OMP) such as multiple angular-domain OMP (MAD-OMP) and polar-domain OMP (PD-OMP)~\cite{cui2023near}. However, these methods incur high computational complexity due to presence of matrix inversions in each iteration, making them unsuitable for real-time applications.
%However, these methods incur high computational complexity due to repeated matrix inversions per iteration and are typically implemented offline, making them unsuitable for real-time applications.
To address this limitation, we develop a NF channel estimation framework based on adaptive filtering, referred to as \textit{polar-domain zero-attracting least mean squares (PD-ZALMS)}. The proposed method achieves significantly superior channel estimation accuracy and lower computational complexity compared with the well-established PD-OMP.
%Moreover, unlike block-based methods that process the entire received pilot signal in each iteration, the proposed PD-ZALMS updates the channel estimate in each iteration using only the pilot signal received up to that point.
Simulation results demonstrate the effectiveness of the proposed algorithm and show that it can even outperform the oracle least-squares (LS) channel estimator at low-to-moderate signal-to-noise ratio (SNR).

\subsubsection*{Notations} Scalars, vectors, and matrices are denoted by lowercase, bold lowercase, and bold uppercase letters, respectively. Transpose, Hermitian transpose, and inverse are represented by $(\cdot)^\tran$, $(\cdot)^\herm$, and $(\cdot)^{-1}$,  respectively. The Hadamard product is denoted by $\odot$. The $n\times n$ identity matrix and the $n \times m$ zero matrix are represented by $ \mathbf{I}_n $ and $\0_{n \times m}$, respectively. The expected value is denoted by $\mathbb{E}\{ \cdot \} $. The gradient operator is represented by $\nabla_{\x} f(\x)$. The sign function is denoted by $\sgn (\cdot)$. The $l_0$-norm and Frobenius norm are represented by $\|\cdot \|_0$ and $\|\cdot \|_{\mathrm{F}}$, respectively.

\section{System Model and Problem Formulation}

This paper focuses on channel estimation in a NF narrowband uplink system, where a base station (BS) equipped with a subarray-based architecture serves $U$ single-antenna users, with $u = 1, \ldots, U$. The BS consists of $M$ subarrays, each with $N$ antennas connected to a radio-frequency (RF) chain via a network of phase shifters, with $m = 1, \ldots, M$ and $n = 1, \ldots, N$. We assume a uniform linear array (ULA) layout with inter-antenna spacing $d$ and inter-subarray spacing $D$. The system model is summarized in Fig.~\ref{fig:mmWave_impr1} (left). Assuming that the users simultaneously transmit orthogonal pilots, the multi-user channel estimation problem can be decomposed into parallel single-user problems, as described next. A summary of the used symbols is given in Table~\ref{table:Nota_tions}.

\subsection{Uplink Training} \label{sec:SM_1}

Let $\H = [\h_{1}, \ldots, \h_{U}] \in \Compl^{MN \times U}$ denote the antenna-domain channel matrix at all the subarrays, where $\h_{u} = [\h_{1,u}^{\tran}, \ldots, \h_{M,u}^{\tran}]^{\tran} \in \mathbb{C}^{M N \times 1}$ is the channel vector between all the subarrays and the $u$-th user, and $\h_{m,u} \in \mathbb{C}^{N \times 1}$ is the channel vector between the $m$-th subarray and the $u$-th user. Furthermore, let $\P = [\p_{1}, \ldots, \p_{U}] \in \mathbb{C}^{Q \times U}$ denote the pilot matrix, where $\p_{u} \in \Compl^{Q \times 1}$ is the pilot of the $u$-th user. Assuming orthogonal pilots, we have $\P^{\herm} \P = Q \I_{U}$. The pre-processing received signal at all the subarrays is given by
\begin{align}
\Y = \H \P^{\herm} + \Z = [\Y_{1}^{\tran}, \ldots, \Y_{M}^{\tran}]^{\tran} \in \Compl^{M N \times Q},
\end{align}
where $\Z = [\Z_{1}^{\tran}, \ldots, \Z_{M}^{\tran}]^{\tran }\in \Compl^{M N \times Q}$ is the additive white Gaussian noise (AWGN) matrix with independent $\setC \setN (0,\sigma^{2})$ elements, and where $\Y_{m} \in \Compl^{N \times Q}$ and $\Z_{m} \in \Compl^{N \times Q}$ are the pre-processing received signal and AWGN matrix, respectively, at the $m$-th subarray.

\begin{table}[t!]
\centering
\label{tab:notations}
\begin{tabular}{|c|c|}
\hline
\textbf{Symbol} & \textbf{Description} \\
\hline
\hline
$M$                             & Number of subarrays (RF chains) \\
$N$                             & Number of antennas per subarray \\
$U$                             & Number of users \\
$Q$                             & Pilot length \\
$\lambda_{\textrm{c}}$          & Carrier wavelength \\
$L$                             & Number of channel paths \\
$\theta_l$                      & Angle of $l$-th path \\
$r_l$                           & Distance of $l$-th path \\
$r^{a(m,n)}$                    & Distance of $n$-th antenna in $m$-th subarray \\
$\h = \G \etab$                 & Antenna-domain channel \\
$\hat{\h}$                      & Estimated antenna-domain channel \\
$\mathbf{G}$                    & Polar-domain dictionary \\
$T_{\theta}$                    & Number of angle samples \\
$T_{r}$                         & Number of distance samples \\
$\etab$                         & Polar-domain channel \\
$\hat{\etab}$                   & Estimated polar-domain channel \\
$\mathbf{W}$                    & Analog sampling matrix \\
$\boldsymbol{\Psi} = \mathbf{W}^\herm \mathbf{G}$ & Sensing matrix \\
\hline
\end{tabular}
\caption{Summary of the used symbols. The angle and distances are with respect to the center of the ULA.}
\label{table:Nota_tions}
\end{table}

To sample each channel $\h_{m,u}$, an analog sampling matrix $\W_{m,u} \in \Compl^{N \times Q}$ is applied at the $m$-th subarray. Then, the pilot contamination from the other users is eliminated by exploiting the pilot orthogonality. The post-processing received signal at the $m$-th subarray for the $u$-th user is given by
\begin{align}
\widetilde{\y}_{m,u} = \W_{m,u}^{\herm} \Y_{m} \p_{u} = \W_{m,u}^{\herm} \h_{m,u} + \widetilde{\z}_{m,u} \in \Compl^{Q \times 1},
\end{align}
with $\widetilde{\z}_{m,u} = \W_{m,u}^{\herm} \Z_{m} \p_{u} \in \Compl^{Q \times 1}$. Finally, the post-processing received signal at all the subarrays for the $u$-th user is obtained as
\begin{align}
\nonumber \widetilde{\y}_{u} & = \W_{u}^{\herm} \Y \p_{u} = \W_{u}^{\herm} \h_{u} + \widetilde{\z}_{u} \\
\label{eq:y_u_tilde} & = [\widetilde{\y}_{m,u}^{\tran}, \ldots, \widetilde{\y}_{M,u}^{\tran}]^{\tran} \in \Compl^{M Q \times 1},
\end{align}
with $\W_{u} = \blkdiag (\W_{1,u}, \ldots, \W_{M,u}) \in \Compl^{M N \times M Q}$ and $\widetilde{\z}_{u} = \W_{u}^{\herm} \Z \p_{u} = [\widetilde{\z}_{1,u}^{\tran}, \ldots, \widetilde{\z}_{M,u}^{\tran}]^{\tran} \in \Compl^{M Q \times 1}$. In this work, we assume that each analog sampling matrix $\mathbf{W}_{m,u}$ is designed as $\mathbf{W}_{m,u} = \mathbf{U} [\mathbf{I}_Q, \mathbf{0}_{Q \times (N-Q)}]^\tran \mathbf{V}^\herm$, where $\mathbf{U}_{m,u} \in \Compl^{N \times N}$ and $\mathbf{V}_{m,u} \in \Compl^{Q \times Q}$ are random unitary matrices, as considered in~\cite{ding2017joint}.

Since the post-processing received signal for the $u$-th user in \eqref{eq:y_u_tilde} is independent of the other users, we omit the user index in the following and consider the channel estimation problem for a generic user based on
\begin{align}\label{rxd_sgn}
\widetilde{\mathbf{y}}=\mathbf{W}^\herm \h+\widetilde{\mathbf{z}} \in \Compl^{M Q \times 1}.
\end{align}

\subsection{Channel Model} \label{sec:SM_2}

We consider a clustered multipath NF channel model~\cite[Ch.~5.6.1]{bjornson2024introduction}. Accordingly, the channel vector of a generic user is given by 
\begin{align}\label{nf_channel}
\h=\sqrt{\frac{{M} N}{L}} \sum_{l=1}^L \gamma_l e^{-j\frac{2\pi}{\lambda_{\textrm{c}}}r_l} \mathbf{g}(\theta_l, r_l),
\end{align}
where $L$ is the number of channel paths, $\gamma_l$ denotes the random complex gain of the $l$-th path, $\theta_l$ and $r_l$ denote the angle and distance, respectively, of the $l$-th path with respect to the center of the ULA, and $\lambda_{\textrm{c}}$ is the carrier wavelength. Furthermore, for a point located at angle $\theta$ and distance $r$, $\mathbf{g}(\theta, r) \in \mathbb{C}^{{M} N \times 1}$ represents the NF array response vector defined as 
\begin{align}
\mathbf{g}(\theta, r) = \frac{1}{\sqrt{M N}} \begin{bmatrix}
e^{-j \frac{2 \pi}{\lambda_{\textrm{c}}} (r^{(a(0,0))} - r)} \vspace{-2mm} \\
\vdots \\
e^{-j \frac{2 \pi}{\lambda_{\textrm{c}}} (r^{(a(0,N-1))} - r)} \vspace{-2mm} \\
\vdots \\
e^{-j \frac{2 \pi}{\lambda_{\textrm{c}}} (r^{(a(M-1,0))} - r)} \vspace{-2mm} \\
\vdots \\
e^{-j \frac{2 \pi}{\lambda_{\textrm{c}}} (r^{(a(M-1,N-1))} - r)} \\
\end{bmatrix},
\end{align}
where $r^{(a(m,n))}$ is the distance from the $n$-th antenna in the $m$-th subarray and $a(m, n)  = (m-1) D+(n-1) d$ is the distance between the $n$-th antenna in the $m$-th subarray and the center of the ULA. With the array deployed along the $z$-axis, as depicted in Fig.~\ref{fig:mmWave_impr1} (right), $r^{(a(m, n))}$ is given by
\begin{align}
r^{(a(m, n))} & = \sqrt{(r \cos (\theta))^2+(a(m, n) - r \sin (\theta))^2} \nonumber\\  & =\sqrt{r^2-2 r a(m, n) \sin (\theta)+a^2(m, n)}.
\end{align}

\subsection{Sparse Channel Representation}
\label{sub_sec_c}
NF channels exhibit sparsity in the polar domain~\cite{cui2022channel}. In this context, let $\mathbf{G} \in \mathbb{C}^{NM \times T_{\theta} T_{r}}$ denote the polar-domain dictionary matrix, where $T_{\theta}$ and $T_{r}$ represent the numbers of angle and distance samples, respectively, with $t_{\theta} = 1, \ldots, T_{\theta}$ and $t_{r} = 1, \ldots, T_{r}$. Hence, the channel in \eqref{nf_channel} can be expressed as
\begin{align}\label{pd_chnl}
\h = \mathbf{G} \etab,
\end{align}
where $\etab \in \mathbb{C}^{T_{\theta} T_{r} \times 1}$ represents the polar-domain channel vector. As discussed in~\cite{cui2023near}, $\mathbf{G}$ comprises $T_{r}$ submatrices, i.e., $\mathbf{G} = [\mathbf{G}_1, \ldots, \mathbf{G}_{T_{r}}]$, with $\mathbf{G}_{t_{r}}=[\mathbf{g}({\theta}_1, {r}_{1, t_{r}} ), \ldots, \mathbf{g}({\theta}_{T_{\theta}}, {r}_{T_{\theta}, t_{r}})] \in \mathbb{C}^{NM\times T_{\theta}}$. The sampled angles and distances must satisfy $\sin {\theta}_{t_{\theta}}=\frac{2(t_{\theta}-1)-T_{\theta}}{T_{\theta}}$ and ${r}_{t_{\theta}, t_{r}}=\frac{D^2 \cos ^2 {\theta}_{t_{\theta}}}{2 \beta^2 \lambda_{\textrm{c}} t_{r}}$, where $\beta$ is a parameter that controls the coherence between the columns of the polar-domain dictionary matrix. By substituting \eqref{pd_chnl} into \eqref{rxd_sgn}, the received signal can be expressed as
\begin{align}\label{rxd_sgn_plr}
\widetilde{\mathbf{y}}=\mathbf{W}^\herm \mathbf{G} \etab+\widetilde{\mathbf{z}} = \boldsymbol{\Psi}\etab + \widetilde{\mathbf{z}},
\end{align}
where $\boldsymbol{\Psi} = \mathbf{W}^\herm \mathbf{G}\in \Compl^{MQ\times T_{\theta}T_{r}}$ is the sensing matrix. Since the number of channel paths satisfies $L\ll T_{\theta}T_{r}$ (especially at high carrier frequencies), the polar-domain channel $\etab$ is sparse and its estimation can be modeled as a sparse signal recovery problem. In this context, our goal is to estimate $\etab$ given $\widetilde{\mathbf{y}}$ and $\boldsymbol{\Psi}$. In the following, we denote the estimated polar-domain channel vector by $\hat{\etab} = [\hat{\eta}_{1}, \ldots, \hat{\eta}_{T_{\theta} T_{r}}] \in \Compl^{T_{\theta} T_{r} \times 1}$ and the corresponding estimated antenna-domain channel vector by $\hat{\h} = \G \hat{\etab} \in \Compl^{M N \times 1}$.

\subsection{Problem Formulation}

The estimation problem for the sparse polar-domain channel
vector $\etab$ can be formulated as
\begin{align}
\begin{array}{cl}
\displaystyle \underset{\etab}{\mathrm{minimize}} & \displaystyle
\|\etab\|_0 \\
\mathrm{s.t.} & \displaystyle \|\widetilde{\mathbf{y}}-\boldsymbol{\Psi}\etab\|_2^2 \leqslant \epsilon,
\end{array}
\end{align}
for a given $\epsilon > 0$. The above optimization problem is inherently non-convex due to the presence of the $l_0$-norm in the objective function. Several sparse signal recovery methods, such as PD-OMP~\cite{cui2022channel} and DL-OMP~\cite{zhang2023near}, have been developed to solve this problem. However, these approaches are computationally intensive, as they involve matrix inversions in each iteration.
%Moreover, they are typically implemented offline, leading to substantial processing delays.
To overcome this limitation, we propose a NF channel estimation framework based on adaptive filtering, termed \textit{polar-domain zero-attracting least mean squares (PD-ZALMS)}, which achieves significantly superior channel estimation accuracy and lower computational complexity compared with PD-OMP.

\section{Proposed PD-ZALMS Algorithm} \label{Section_iii}

In this section, we first outline the core principles of adaptive filtering for channel estimation and then introduce the proposed PD-ZALMS algorithm. Unlike the conventional zero-attracting least mean squares methods, PD-ZALMS operates in the polar domain, enabling efficient channel estimation in NF XL-MIMO systems.

\subsection{Adaptive Filtering for Channel Estimation}

Adaptive filtering algorithms operate through two fundamental stages:
\begin{enumerate}
    \item \textit{Filtering stage}, which includes: a) calculating the output of a finite impulse response filter (FIR) based on a series of tap inputs, and b) determining the estimation error by comparing the filter's output with a target signal.
    \item \textit{Adaptation phase}, which involves automatically updating the filter's tap weights based on the estimation error.
\end{enumerate}
The integration of these two stages forms a feedback loop around the adaptive filtering algorithm, as depicted in Fig.~\ref{fig:adative_filter}. The observation error at the $k$-th iteration is defined as
\begin{align}\label{err_eq}
\mathbf{e}(k) = \widetilde{\y} - \boldsymbol{\Psi}\hat{\etab}(k-1) \in \Compl^{M Q \times 1},
\end{align}
where  $\hat{\etab}(k-1)$ is the estimated channel at the $(k-1)$-th iteration. The adaptive filter processes the input signal to produce the output $\boldsymbol{\Psi}\hat{\etab}(k-1)$, typically using a FIR structure. The observation error $\mathbf{e}(k)$ is computed as the difference between the received and output signals. The adaptive update process computes the gradient vector based on the observation error $\mathbf{e}(k)$ and updates the filter weights $\hat{\etab}(k)$ by minimizing the mean squared error (MSE) cost function
\begin{align}
J(k) = \mathbb{E}\{\|\widetilde{\y} - \boldsymbol{\Psi}\hat{\etab}(k-1)\|^2\},
\end{align}
where the expectation is taken over $\W$, $\tilde{\z}$, and $\{ \gamma_{l} \}_{l=1}^{L}$. The minimization of the cost function is performed through iterative updates along the direction of the negative gradient as~\cite{elangovan2012comparative}
\begin{align}
\hat{\etab}(k)=\hat{\etab}(k-1)-\frac{\mu}{2} \nabla_{\hat{\etab}(k-1)}({{J}}(k)),
\end{align}
where the step size $\mu$ plays a critical role in determining the convergence rate. Cost functions vary across adaptive filtering algorithms, and algorithm selection significantly impacts convergence rate and estimation accuracy. Nevertheless, these algorithms share a common goal, that is, to find an estimate $\hat{\etab}$ of the true channel $\etab$.

\tikzstyle{arrow} = [thin,->,>=stealth]
	{\begin{figure}	
			\begin{tikzpicture}	[scale=.7,transform shape ]
				
		% Draw rectangle--------	
\draw [rounded corners=5pt, fill=gray!10!white,draw=green, thick]
				(.5,-.5) rectangle (11.8,-4.5) ;

	\draw [rounded corners=5pt, fill=gray!30!white]
				(2,0) rectangle (4,1) ;
				\node at (3,.5) { $ \etab $};
%		\node at (3,.6) { \textcolor{blue}{True channel}};
        \node at (5,.8) {$\boldsymbol{\Psi}\etab$};
		\draw [arrow] (4.5,-3.5)--(2,-.75);

				\draw [rounded corners=5pt, fill=gray!30!white]
				(2,-1) rectangle (4,-2) ;
\node at (5,-1.2){$\boldsymbol{\Psi}\hat{\etab}(k-1)$};

\node at (3,-1.5) { $  \hat{\etab}(k-1) $};
%\node at (3,-1.2) {\textcolor{blue}{Estimated}};	
%	\node at (3,-1.5) { \textcolor{blue}{channel}};	
		\draw [rounded corners=5pt, fill=gray!30!white]
			(5,-3) rectangle (7,-4);
			
	\node at (6,-3.3) { \textcolor{blue}{Adaptive} };
	\node at (6,-3.7) { \textcolor{blue}{algorithm}};

% Draw first circle------------ 		
\draw [black, fill=yellow] (7,.5) circle (.5);
\node at (6,0.8) {\huge $+$};
\node at (6.6,1.2) {\huge $+$};

% Draw second circle------------ 	
\draw [black, fill=yellow] (10,-1.5) circle (.5);
\node at (9,-1.2) {\huge $-$};
\node at (9.6,-.8) {\huge $+$};

% Write variables--------------------
	
\node at (.1,.5) {$\boldsymbol{\Psi}$};	
\node at (11.4,.5) {$\widetilde{\y}$};
%\node at (9,.8) {Desired Signal};
\node at (7.4,1.25) {$\widetilde{\mathbf{z}}$};
\node at (11.4,-1.5) {$\e(k)$};
	
% Draw arrow-------------------	
				
\draw [arrow] (.5,.5)--(2,.5);
\draw [arrow] (4,.5)--(6.5,.5);
\draw [arrow] (7.5,.5)--(11,.5);
\draw [arrow] (10,.5)--(10,-1);
\draw [arrow] (7,1.5)--(7,1);

\draw [arrow] (1,-1.5)--(2,-1.5);

\draw [arrow] (4,-1.5)--(9.5,-1.5);

\draw [arrow] (10.75,-3.5)--(7,-3.5);
\draw [arrow] (10.5,-1.5)--(11,-1.5);

% Draw line-------------------
				
	\draw [line width=.5] (1,.5)--(1,-1.5);				
				
\draw [line width=.5] (10.75,-1.5)--(10.75,-3.5);

\draw [line width=.5] (4.5,-3.5)--(5,-3.5);

			\end{tikzpicture}
			\caption{Block diagram of channel estimation via adaptive filtering.}
			\label{fig:adative_filter}
\end{figure}
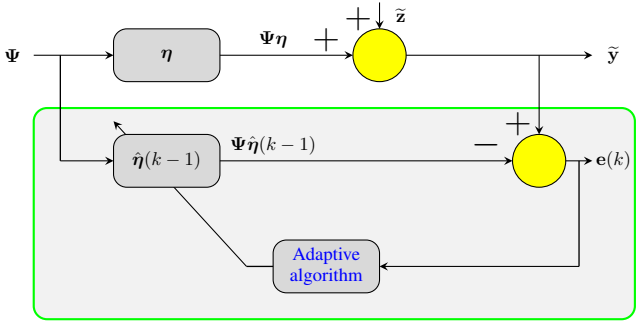}

\subsection{Proposed Method}

To estimate the sparse polar-domain channel vector at the $k$-th iteration, the proposed PD-ZALMS algorithm utilizes a regularized cost function that combines the MSE and a penalty term~\cite{jin2010stochastic, chen2010regularized}. Accordingly, the cost function can be expressed as
\begin{align}
{\mathcal{J}}(k) = \mathbb{E}\{\|\widetilde{\y} - \boldsymbol{\Psi}\hat{\etab}(k-1)\|^2\} + \delta^{\prime} f(\hat{\etab}(k-1)),
 \end{align}
where $f(\cdot)$ is a sparsity-inducing penalty function and $\delta^{\prime}$ is a regularization parameter. Using gradient descent~\cite{haykin2002adaptive}, the channel estimate is iteratively updated as
\begin{align}\label{wt_up}
\hat{\etab}(k) = \hat{\etab}(k-1) - \frac{\mu}{2} \nabla_{\hat{\etab}(k-1)}({\mathcal{J}}(k)),
\end{align}
where the step size $\mu$ controls the convergence rate and error floor. The gradient of the cost function is given by
\begin{align}
\nonumber \nabla_{\hat{\etab}(k-1)}(\mathcal{J}(k)) = \ & 2 \mathbf{R} \hat{\etab}(k-1) - 2 \mathbf{c} \\
\label{GD_Eq} & - \delta \nabla_{\hat{\etab}(k-1)}(f(\hat{\etab}(k-1))),
\end{align}
where $\mathbf{R} = \mathbb{E}\{\boldsymbol{\Psi}^{\mathrm{H}}\boldsymbol{\Psi}\}\in \Compl^{T_{\theta}T_{r}\times T_{\theta}T_{r}}$ is the autocorrelation matrix of the sensing matrix, $\mathbf{c} = \mathbb{E}\{\boldsymbol{\Psi}^{\mathrm{H}}\widetilde{\mathbf{y}}\}\in \Compl^{T_{\theta}T_{r}\times 1}$ is the cross-correlation vector, and $\delta = \frac{\mu \delta^{\prime}}{2}$ is the regularization step size. Substituting \eqref{GD_Eq} into \eqref{wt_up} yields
\begin{align}
\nonumber \hat{\etab}(k) = \ & \hat{\etab}(k-1) - \mu \bigg( \mathbf{R} \hat{\etab}(k-1) - \mathbf{c} \\
\label{upd_eq} & - \frac{\delta}{2} \nabla_{\hat{\etab}(k-1)}(f(\hat{\etab}(k-1))) \bigg).
\end{align}
Accurately estimating the parameters $\mathbf{R}$ and $\mathbf{c}$ is challenging, as their estimation requires averaging over many samples. To address this, the proposed PD-ZALMS algorithm employs a stochastic gradient approach~\cite{haykin2002adaptive}, replacing $\mathbf{R}$ and $\mathbf{c}$ with their instantaneous approximations $\boldsymbol{\Psi}^{\mathrm{H}}\boldsymbol{\Psi}$ and $\boldsymbol{\Psi}^{\mathrm{H}}\widetilde{\y}$, respectively. After rearranging the terms, \eqref{upd_eq} can be simplifies as
\begin{align}\label{upd_eq22}
\hat{\etab}(k) = \hat{\etab}(k-1) + \mu \boldsymbol{\Psi}^{\mathrm{H}}\mathbf{e}(k) - \delta \nabla_{\hat{\etab}(k-1)}(f(\hat{\etab}(k-1))).
\end{align}

In this work, we use the $l_0$-norm as sparsity-inducing penalty function, as it yields the sparsest channel estimate. Hence, we have
\begin{align}
f(\hat{\etab})=\|\hat{\etab}\|_0=\sum_{i=1}^{T_{\theta} T_{r}} \mathcal{I} (|\hat{{\eta}_i}|),
\end{align}
where $\mathcal{I}(\cdot)$ denotes the indicator function that is equal to $0$ if the argument is zero and to $1$ otherwise. Since the $l_0$-norm is non-convex and non-differentiable, complicating gradient-based optimization, we adopt a tractable approximation similar to those described in~\cite{jin2010stochastic, chen2010regularized}, i.e.,
\begin{align}\label{appr_eqa}
\sum_{i=1}^{T_{\theta} T_{r}} \mathcal{I}(|\hat{{\eta}}_i|) \approx \sum_{i=1}^{T_{\theta} T_{r}}(1-e^{-\alpha |\hat{{\eta}}_i|}),
\end{align}
where $\alpha$ is a parameter that defines the accuracy of the approximation. Thus, the gradient   of the penalty function after the approximation is obtained as
\begin{align}\label{appr_norm}
\nabla_{\hat{\etab}(k-1)}(f(\hat{\etab}(k-1))) =  \alpha e^{-\alpha |\hat{\etab}(k-1)|} \odot \sgn(\hat{\etab}(k-1)),
\end{align}
where $e^{-\alpha |\hat{\etab}(k-1)|}$ denotes the element-wise exponential of $-\alpha |\hat{\etab}(k-1)|$ and $|\hat{\etab}(k-1)|$ is the element-wise absolute value of $\hat{\etab}(k-1)$. Finally, substituting \eqref{appr_norm} into \eqref{upd_eq22} yields the update equation for the PD-ZALMS algorithm, given by 
\begin{align}
\nonumber \hat{\etab}(k) = \ & \hat{\etab}(k-1)  +\mu \boldsymbol{\Psi}^\herm\mathbf{e}(k) \\
\label{upd_equ3} & -\delta\alpha e^{-\alpha |\hat{\etab}(k-1)|} \odot \sgn(\hat{\etab}(k-1)),
\end{align}
where $\delta\alpha$ balances sparsity and estimation accuracy. 

The initialization and step-by-step process for the proposed PD-ZALMS method are outlined in Algorithm~\ref{PD_ZALMS}. In step 1, we initialize $\hat{\etab}(0) = \mathbf{0}_{T_{\theta}T_{r}\times 1}$. Next, in steps 2--4, the sensing matrix $\boldsymbol{\Psi}$ is constructed. Then, in steps 6--7, the observation error and the update filter weights are calculated. Finally, the estimated value of the antenna-domain NF channel is obtained in step 9.

\begin{algorithm}[t!]
\caption{Proposed PD-ZALMS Algorithm}\label{PD_ZALMS}
\textbf{Input:}  $\widetilde{\mathbf{y}}$, $\boldsymbol{\Psi}$, $\mu$, $\delta$, and $\alpha$.\\ 
	\textbf{Output:} $\hat\h$.
	\begin{algorithmic}[1]
		\State \textbf{Initialization:} $\hat{\etab}(0) = \mathbf{0}_{T_{\theta}T_{r}\times 1}$.
        \State Construct the analog sampling matrix $\mathbf{W}$ (see Section~\ref{sec:SM_1}).
        \State Construct the polar-domain dictionary matrix $\mathbf{G}$ (see Section~\ref{sub_sec_c}). 
        \State Compute the sensing matrix $\boldsymbol{\Psi} = \mathbf{W}^\herm \mathbf{G}$.
		\State \textbf{for} $k = 1, 2, \ldots$: \textbf{Do}
		\State \quad Compute $\mathbf{e}(k)$ as in \eqref{err_eq}.
		\State \quad Update $\hat{\etab}(k)$ as in \eqref{upd_equ3}.
		\State\textbf{end}
		\State\textbf{Return} $\hat\h = \mathbf{G}\hat{\etab}$.
	\end{algorithmic}
\end{algorithm}

\subsection{Computational Complexity} \label{paper_complexity_ana}

The computational complexity of the proposed PD-ZALMS method in each iteration is reported in Table~\ref{table:computation_compl} along with that of the baseline schemes MAD-OMP, PD-OMP, and oracle LS. The complexity of MAD-OMP is lower than that of PD-OMP since MAD-OMP relies solely on angular-domain sampling, yielding a complexity of $\mathcal{O}(MQ T_{\theta} + (M T_{\theta})^3)$. On the other hand, PD-OMP requires both angular- and distance-domain sampling, resulting in a higher complexity of $\mathcal{O}(M Q T_{\theta} T_{r} + (M T_{\theta} T_{r})^3)$. The complexity of the proposed scheme is specifically determined by steps 6--7 in Algorithm~\ref{PD_ZALMS}. Step 6 involves the multiplication of $\boldsymbol{\Psi} \in \mathbb{C}^{MQ \times T_{\theta}T_{r}}$ with $\hat{\etab} \in \mathbb{C}^{T_{\theta}T_{r} \times 1}$ followed by the addition of $\widetilde{\mathbf{y}}$, resulting in a complexity of $\mathcal{O}(MQ T_{\theta}T_{r})$ in each iteration. Step 7 involves the multiplication of $\boldsymbol{\Psi}^\textrm{H}(k)$ with $\mathbf{e}(k)$, which incurs a complexity of $\mathcal{O}(MQ T_{\theta}T_{r})$. Hence, the total computational complexity of the proposed scheme is $\mathcal{O}(2MQ T_{\theta}T_{r})$ in each iteration. Notably, the proposed scheme avoids matrix inversion, leading to a significantly lower computational burden compared with the OMP-based baseline methods that require matrix inversion in each iteration.

\begin{table}[t!]
	\begin{center}
		\begin{tabular}{ |c|c| } 
			\hline
			\textbf{Algorithm} & \textbf{Computational complexity} \\ 
			\hline
            \hline
			MAD-OMP &  $\mathcal{O}(MQT_{\theta} + (MT_{\theta})^3)$\\
			PD-OMP 	& $\mathcal{O}(MQT_{\theta}T_{r} + (MT_{\theta}T_{r})^3)$	 \\
			Oracle LS	& $\mathcal{O}(2MQL^2+L^3+MT_{\theta}T_{r}L)$\\
            PD-ZALMS (proposed) &  $\mathcal{O}(2MQT_{\theta}T_{r})$\\
			\hline
		\end{tabular}
		
	\end{center}
    \caption{Computational complexity of the proposed and baseline algorithms.} 
    \label{table:computation_compl}
\end{table}

\section{Numerical Results} \label{sec:Simulation Results}

In this section, we present simulation results to demonstrate the performance of the proposed PD-ZALMS algorithm for a narrowband sub-THz XL-MIMO system. The simulation parameters are listed in Table~\ref{table:simulation_par}, which consider a carrier frequency of $100$~GHz. The inter-antenna spacing is $d = \frac{\lambda_{\text{c}}}{2} = 1.5~\text{mm}$, the inter-subarray spacing is $D = Nd + 8\lambda_{\text{c}} = 72~\text{mm}$, the ULA aperture is $MD = M(Nd + 8\lambda_{\text{c}}) = 0.576~\text{m}$, and the Fraunhofer distance is $R_{\textrm{FD}} = \frac{2 (MD)^2}{\lambda_{\text{c}}} = 221~\text{m}$. We consider a generic user whose channel consists of $L = 4$ paths with uniformly distributed angles and distances, where the angles satisfy $\sin\theta \in [-0.75, 0.75]$ and the distances vary from $5$~m to $R_{\textrm{FD}}$. The grid is defined by the sine of the angle with an angular resolution of $2/N$ and a distance resolution of $5$~m. The complex path gain of the $l$-th path is distributed as $\gamma_l \sim \mathcal{CN}(0,1)$, $\forall l$, as in~\cite{cui2022channel}. As a performance metric, we consider the normalized mean square error (NMSE) defined as $\text{NMSE} = \mathbb{E}\{\|\h - \hat{\h}\|_2^2 / \|\h\|_2^2\}$. The oracle LS method, which assumes knowledge of the polar-domain channel support, is also included for comparison. The plots are obtained through Monte Carlo simulations with $2000$ independent channel realizations and all the methods are evaluated after convergence.

\begin{table}[t!]
\label{table:1}	
	\begin{center}
		\begin{tabular}{ |c|c| } 
			\hline			
			\textbf{Simulation parameter} & \textbf{Value} \\ 
            \hline
			\hline
			Number of subarrays & $M = 8$ \\
            Number of antennas per subarray &$N = 32$ \\
            Carrier wavelength & $\lambda_{\textrm{c}} = 3$~mm \\
            Inter-antenna spacing & $d = \lambda_{\textrm{c}}/2$ \\
            Inter-subarray spacing & $D = Nd + 8 \lambda_{\textrm{c}}$ \\
            Number of channel paths & $L = 4$ \\
			Step size & $\mu = 6\times 10^{-5}$ \\
            Regularization step size &	$ \delta = 5 \times 10^{-5} $ \\
			Accuracy parameter & $\alpha = 25$ \\
			\hline
		\end{tabular}
	\end{center}
    
    \caption{Simulation parameters.} 
    \label{table:simulation_par}
\end{table}

Fig.~\ref{fig:nmse_vs_snr} depicts the NMSE versus the SNR, with pilot length $Q = 15$. As the SNR increases, the NMSE performance of all the algorithms improves. PD-OMP always outperforms MAD-OMP, as it uses a polar-domain dictionary based on spherical wavefronts, whereas MAD-OMP utilizes an angular-domain dictionary based on planar wavefronts~\cite{cui2022channel, zhang2023near}. The proposed PD-ZALMS algorithm significantly surpasses both OMP-based schemes. At low SNR (from $-10$ to $0$ dB), PD-ZALMS performs slightly better than PD-OMP. However, at higher SNR (above $0$~dB), PD-ZALMS significantly outperforms both MAD-OMP and PD-OMP: for example, at $30$~dB, it achieves an NMSE approximately $15$~dB lower than that of PD-OMP. In the SNR range between $6$ and $25$~dB, it even outperforms the oracle LS estimator, which assumes knowledge of the polar-domain channel support. Additionally, PD-ZALMS has significantly lower computational complexity, as discussed in Section~\ref{paper_complexity_ana}.

\begin{figure}[t!]
\centering
\begin{tikzpicture}

\begin{axis}[
	width=8cm, height=7cm,
	xmin=3, xmax=30,
	ymin=-35, ymax=5,
	xlabel={Pilot length},
	ylabel={NMSE [dB]},
	xlabel near ticks,
	ylabel near ticks,
	label style={font=\footnotesize},
	xtick={3,6,...,30},
	ytick={-35,-30,...,5},
%	tick scale binop=\times,
%	xticklabels={},
%	yticklabels={},
    ticklabel style={font=\footnotesize},
	legend style={at={(0.98,0.8)}, anchor=north east, font=\scriptsize, inner sep=1pt, fill opacity=0.75, draw opacity=1, text opacity=1},
	legend cell align=left,
	title style={font=\scriptsize, yshift=-2mm},
	grid=major
]

\addplot[thick, blue, mark=+]
table [x=Pilot, y=MADOMP, col sep=comma] {Results/files_txt/NMSE_Pilot.txt};
\addlegendentry{MAD-OMP};

\addplot[thick, red, mark=square]
table [x=Pilot, y=PDOMP, col sep=comma] {Results/files_txt/NMSE_Pilot.txt};
\addlegendentry{PD-OMP};

\addplot[thick, black, mark=o]
table [x=Pilot, y=OLS,, col sep=comma] {Results/files_txt/NMSE_Pilot.txt};
\addlegendentry{Oracle LS};

\addplot[thick, green, mark=triangle]
table [x=Pilot, y=PDZALMS,, col sep=comma] {Results/files_txt/NMSE_Pilot.txt};
\addlegendentry{PD-ZALMS (proposed)};

\end{axis}

\end{tikzpicture}
\caption{NMSE versus pilot length of MAD-OMP, PD-OMP, oracle LS, and the proposed PD-ZALMS, with $\textrm{SNR} = 15$~dB.}
\label{fig:nmse_vs_pilot}
\end{figure}
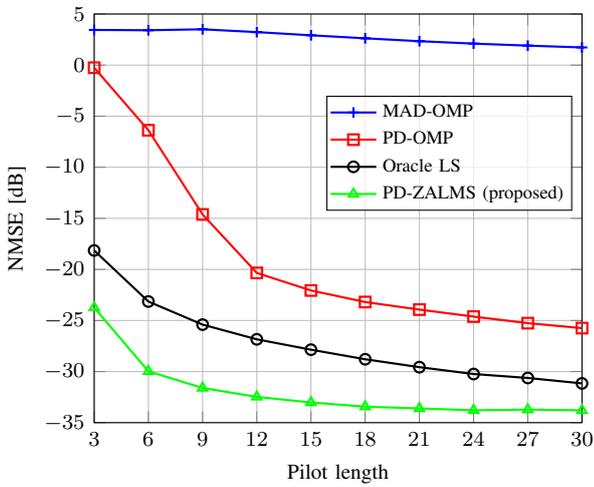

Fig.~\ref{fig:nmse_vs_pilot} shows the NMSE versus the pilot length, with $\textrm{SNR}=15$~dB. As expected, the NMSE performance of all the schemes improves with the pilot length and that the proposed PD-ZALMS algorithm consistently outperforms all the baseline methods. For example, it achieves an NMSE gain over PD-OMP of about $11$~dB with $Q = 15$ and of about $3$~dB with $Q = 30$. These results highlight that PD-ZALMS can achieve substantially better performance or significant pilot overhead reduction with respect to well-established alternatives.

\begin{figure}[t!]
\centering
\begin{tikzpicture}

\begin{axis}[
	width=8cm, height=7cm,
	xmin=-15, xmax=40,
	ymin=-60, ymax=20,
	xlabel={SNR [dB]},
	ylabel={NMSE [dB]},
	xlabel near ticks,
	ylabel near ticks,
	label style={font=\footnotesize},
	xtick={-15,-10,...,40},
	ytick={-60,-50,...,40},
%	tick scale binop=\times,
%	xticklabels={},
%	yticklabels={},
    ticklabel style={font=\footnotesize},
	legend style={at={(0.02,0.02)}, anchor=south west, font=\scriptsize, inner sep=1pt, fill opacity=0.75, draw opacity=1, text opacity=1},
	legend cell align=left,
	title style={font=\scriptsize, yshift=-2mm},
	grid=major
]
\addplot[thick, blue, mark=+]
table [x=SNR, y=MADOMP, col sep=comma] {Results/files_txt/NMSE_SNR.txt};
\addlegendentry{MAD-OMP};
\addplot[thick, red, mark=square]
table [x=SNR, y=PDOMP, col sep=comma] {Results/files_txt/NMSE_SNR.txt};
\addlegendentry{PD-OMP};
\addplot[thick, black, mark=o]
table [x=SNR, y=OLS,, col sep=comma] {Results/files_txt/NMSE_SNR.txt};
\addlegendentry{Oracle LS};
\addplot[thick, green, mark=triangle]
table [x=SNR, y=PDZALMS,, col sep=comma] {Results/files_txt/NMSE_SNR.txt};
\addlegendentry{PD-ZALMS (proposed)};

\end{axis}

\end{tikzpicture}
\caption{NMSE versus SNR of MAD-OMP, PD-OMP, oracle LS, and the proposed PD-ZALMS, with $Q = 15$. }
\label{fig:nmse_vs_snr}
\end{figure}
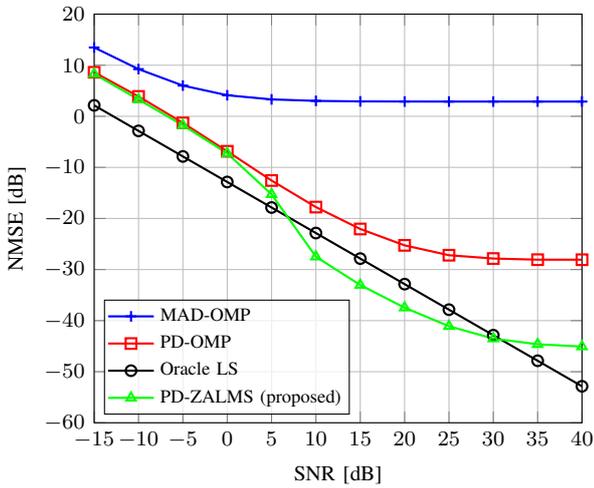

\section{Conclusions} \label{sec:Conclusion}

%In this paper, we investigate channel estimation in a sub-THz XL-MIMO system, considering NF channel properties. Leveraging polar-domain sparsity, we propose the PD-ZALMS algorithm, based on adaptive filtering. Simulation results show that the PD-ZALMS algorithm outperforms existing algorithms, such as PD-OMP, MAD-OMP, and even oracle LS for certain SNR range in estimation accuracy and pilot efficiency. Thus, the PD-ZALMS algorithm is the preferred choice for NF sub-THz XL-MIMO systems due to its superior accuracy and efficiency. In this paper, we have considered the on-grid parameter, however, for better estimation accuracy, the off-grid parameter will be considered in future works.
This work addressed sparse channel estimation for XL-MIMO systems in the NF regime. Assuming a practical subarray-based architecture, we developed the PD-ZALMS algorithm, a NF channel estimation framework based on adaptive filtering. The proposed method achieves significantly superior channel estimation accuracy and lower computational complexity compared with the well-established PD-OMP algorithm, and can even outperform the oracle LS channel estimator at low-to-moderate SNR. The proposed PD-ZALMS algorithm can be extended to operate in a fully online manner, such that, at each iteration, only the signal received up to that point is processed instead of waiting for the entire pilot signal: this will be addressed in future work.

\newpage

\bibliographystyle{IEEEtran}
\bibliography{main_final,refs_abbr}
\end{document}